\documentstyle[aps,prd,floats]{revtex}

\preprint{DAMTP-2000-49}
\date{January 16, 2001}
\tighten

\begin{document}
\draft
\newcommand\lsim{\mathrel{\rlap{\lower4pt\hbox{\hskip1pt$\sim$}}
    \raise1pt\hbox{$<$}}}
\newcommand\gsim{\mathrel{\rlap{\lower4pt\hbox{\hskip1pt$\sim$}}
    \raise1pt\hbox{$>$}}}


\title{Primordial Gaussian Fluctuations from Cosmic Defects}

\author{P. P. Avelino${}^{1,2}$\thanks{
Electronic address: pedro\,@\,astro.up.pt} and
C. J. A. P. Martins${}^{1,3}$\thanks{
Electronic address: C.J.A.P.Martins\,@\,damtp.cam.ac.uk}
}

\address{${}^1$ Centro de Astrof\'{\i}sica, Universidade do Porto\\
Rua das Estrelas s/n, 4150-762 Porto, Portugal}

\address{${}^2$ Dep. de F{\' \i}sica da Faculdade de Ci\^encias da 
Univ. do Porto\\
Rua do Campo Alegre 687, 4169-007 Porto, Portugal}

\address{${}^3$ Department of Applied Mathematics and Theoretical Physics\\
Centre for Mathematical Sciences, University of Cambridge\\
Wilberforce Road, Cambridge CB3 0WA, U.K.}

\maketitle
\begin{abstract}

{We extend our recent work on ``two-metric'' theories of
gravity by showing how in such models cosmic defects can produce a spectrum
of primordial Gaussian density perturbations. This will happen when the
speed characterising the decay products of the defect network is much larger
than the speed characterising gravity and all standard model particles.
This model will exactly mimic all the standard predictions of inflationary
models, and the only way of distinguishing the two will be via the detection
of the decay products of the network.}

\end{abstract} 

\section{Introduction}
\label{secintro} 

The recent high-resolution measurements of Cosmic Microwave Background (CMB)
anisotropies on degree and sub-degree scales \cite{boo1}
have started to provide
cosmologists with a realistic chance of probing the main physical mechanisms
and conditions of the early universe \cite{kolb}. However, one too often
forgets that any subsequent treatment of the data (eg, to determine
``preferred'' cosmological parameters) will require model-dependent
assumptions, and one must make sure that these are properly justified.
Two recent papers \cite{boo2} illustrate a rather eclectic range of
ways in which to use a given data set.

Consider the two basic paradigms that could be responsible
for producing these anisotropies---topological defect \cite{vsh} and
inflationary \cite{linde} models. There are of course some rather generic
differences between the two, but still the task
of unambiguously distinguishing between them is not at
all trivial \cite{hor,liddle}.
The presence of super-horizon perturbations or of ``Doppler peaks'' \cite{gun}
on small angular scales, for example, are not good discriminants \cite{turok}
(at least if they are taken on their own).
This issue is further complicated since one can easily
obtain `natural' models where both defects and
inflation generate density fluctuations \cite{acm2,cb}.

In previous work \cite{previous}, we have presented an explicit example of
a mechanism whereby the primordial fluctuations are generated
by a network of cosmic defects, but are nevertheless very similar to a
standard inflationary model. Such models arise in the context
of ``two-metric'' theories \cite{mof,dru,bass}.
The only difference between the observational consequences of these models
and those of the standard inflationary
scenario is a relatively small non-Gaussian component.
Here we discuss an alternative model arising in the same context,
but where the defect-induced primordial fluctuations are also Gaussian.
In this case the evolution of the defect network is standard,
but we require that the characteristic speed of the decay products of
the defect network is much larger than the speed characterizing gravity
and all the standard model particles. We show that this model
will exactly reproduce the CMB and large-scale structure (LSS)
predictions of the standard inflationary models,
and the only way to identify it would be through the decay products of
the defect network involved.

\section{The model}
\label{model} 

As in our previous work \cite{previous},
we shall consider the so-called ``two-metric'' theories \cite{mof,dru,bass},
which contain two natural speed parameters.
In \cite{previous} we assumed that the scalar field which produced the defects
had a characteristic speed $c_\phi$ that was much larger than
the characteristic speed of gravity and the standard model interactions.
Here we shall instead concentrate on the decay products of the defect
network.

In recent years there has been some debate over the issue of which is the
dominant energy loss mechanism for defect networks. In particular, for
the case of cosmic strings, there are two quite distinct possibilities,
namely an energy loss mechanism based on gravitational radiation \cite{vsh}
and one based on particle production \cite{vah,moore}. Furthermore, some
preliminary studies \cite{chmold,rdp} indicate that the cosmological
consequences---say, as measured from the CMB and matter power spectra---have
some significant differences, although one should be cautious given the
number of simplifying assumptions made in order to derive these results.

In this paper we shall assume a two-metric theory where cosmic strings are
produced, and where the long string network evolves in the standard
way (with the comoving correlation length of the network being of the order 
of $c \eta$), and suppose that this network decays via particle production.
Furthermore, we also assume that the decay products have a characteristic
speed $c_p$ which obeys $c_p \gg c$. In this case, the point will be to 
have  the ``compensation'' scale much larger than the horizon
size (defined in the usual way)---note that this will be so because
one expects the compensation scale to be of the order of the free
streaming length of the decay products.

As in \cite{previous}, the point of this paper is to describe a rather
simple and general mechanism and its basic consequences, rather than a
specific realization of it. Hence we will not concern ourselves
with discussing
particular models. In particular, the mechanism we will be discussing is
independent of $c_p$ being a constant or a time-varying quantity. 
For the time being we shall assume that $c_p$ is a time-independent 
quantity, simply for the purpose of simplifying the discussion.
We will later relax this assumption and discuss the implications 
for structure formation of variable $c_p$.
Moreover,
what the ultimate decay products of the defect network {\em are} will also not
affect the model's ability to work, although it will of course affect its
detailed quantitative predictions (or, if one wants to put it in other words,
its efficiency).

For example, the decay products could be massless particles with velocity
$c_p \gg c$; these would, in analogy with the Cerenkov effect, emit
gravitons and thus produce a stochastic gravitational wave
background \cite{bass}. On the other hand, if the decay products are
massive particles produced with characteristic velocity $c_p$, then we
expect them to eventually be slowed down to below the characteristic
speed of gravity, on a timescale
related to the graviton emission rate. As will become clear below, our
mechanism will be more efficient in the former case than in the latter.
On the other hand, the subsequent evolution of these decay products can
potentially be used to impose constraints on specific realizations of the
mechanism, and the detection of a ``background'' of these decay products
would provide an obvious way to test the mechanism.

The only detailed assumption we require is that
the decay of the defect network
proceeds in such a way to allow its evolution
to be {\em qualitatively analogous}
to the standard case \cite{vsh,ms1,ms2,thesis}. In particular, we assume that
some ``scaling'' solution will be reached
after a relatively short transient period. However, we do allow (and indeed
expect) a ``scaling'' solution that is {\em quantitatively different}
from the usual, ``scale-invariant'' one---see \cite{model} for a more
detailed discussion of these concepts. This assumption still allows us to
make use of some of the standard results on string-seeded structure
formation in the following section.

We emphasize that there are significant differences between the model
being discussed here and the one previously presented in \cite{previous}.
In our previous work the speed characterizing the defect-producing
scalar field $c_\phi$ was much larger than the speed $c$ characterizing
gravity and standard model particles. Hence the comoving correlation
length of the network was of order $c_\phi \eta \gg c \eta$. On the
other hand, the decay products of the network were assumed to be standard,
ie to have a characteristic speed $c$. In the present work we assume
that the evolution of the network is standard (with the comoving correlation 
length of the network being of the order of $c \eta$), but that 
its decay mechanism is through a channel with some characteristic speed 
$c_p$ that is much larger than the standard one.

\section{Cosmological consequences}
\label{growth}

In the synchronous gauge, the linear evolution equations for radiation and 
cold dark matter perturbations, $\delta_r$ and $\delta_m$, in a 
flat universe with zero cosmological constant are
\begin{eqnarray}
  \ddot \delta_m + {\dot a \over a} \dot \delta_m -
  {3 \over 2}\Big({\dot a \over a}\Big)^2 \, \left({a
      \delta_m + 2 a_{eq} \delta_r
      \over a + a_{eq}}\right)  = 4 \pi G 
  \Theta_+,
  \label{one}\\
  \ddot \delta_r - {1 \over 3} \nabla^2 \delta_r
  - {4 \over 3}
  \ddot \delta_m = 0\, ,
  \label{two}
\end{eqnarray}
where $\Theta_{\alpha \beta}$ is the energy-momentum
tensor of the external source, $\Theta_+ =  \Theta_{00} +\Theta_{ii}$, $a$
is the scale factor,
``{\it eq}'' denotes the epoch of radiation-matter equality,
and a dot represents a derivative with
respect to conformal time. We will consider the growth of super-horizon 
perturbations with $c k \eta \ll 1$. Then eqn. (\ref{one}) becomes:
\begin{equation}
 \ddot \delta_m + {\dot a \over a} \dot \delta_m -
  {1 \over 2}\Big({\dot a \over a}\Big)^2 \, \left({3a
      + 8 a_{eq} 
      \over {a + a_{eq}}}\right) \delta_m  = 4 \pi G 
  \Theta_+\, ,
  \label{three}
\end{equation}
and $\delta_r=4\delta_m/3$.
Its solution, with initial conditions 
$\delta_m =0$, ${\dot \delta_m}=0$ can be written as
\begin{equation}
\delta^S_{m}({\bf x},\eta) = 4 \pi G \int_{\eta_i}^\eta d\eta' \,
\int d^3x' {\cal G}(X;\eta,\eta') \Theta_{+}({\bf x'},\eta')\, ,
\end{equation}
\begin{equation}
{\cal G}(X;\eta,\eta') = {1 \over 2 \pi^2} \int_0^\infty \, \widetilde {\cal
G}(k;\eta,\eta') {\sin k X \over k X} k^2 dk\, . 
\end{equation} 
Here $X=|{\bf x} -{\bf x'}|$ and `S'
indicates that these are the `subsequent' fluctuations, according to the
notation of \cite{VS}, to be distinguished from `initial' ones.

We are interested in computing the inhomogeneities at late times in
the matter era. When
$\eta_0 \gg \eta_{eq}$, the Green functions are dominated by the growing mode,
$\propto a_0/a_{eq}$, so the function we would like to solve for is \cite{VS}
\begin{equation}
T(k;\eta) = \lim_{\eta_0/\eta_{eq} \to \infty} {a_{eq} \over a_0}
\widetilde {\cal G}(k,\eta_0,\eta)\, .
\label{transfer}
\end{equation}
Consider the growth of super-horizon perturbations generated during 
the radiation era, for which the transfer function can be
written\footnote{For super-horizon perturbations generated during
the matter era, the transfer function would differ from the above
by a factor of two.} \cite{VS}
\begin{equation}
T(0;\eta) = \frac{\eta_{eq}}{10(3-2{\sqrt 2}) \eta}\, .
\label{T0}
\end{equation}
The linear perturbations induced by defects such as cosmic strings, are
the sum of initial and subsequent perturbations:
\begin{eqnarray}
\delta_m(k;\eta_0) &=& \delta_m^I(k;\eta_0) + \delta_m^S(k;\eta_0)
\cr\cr &=& 4 \pi G (1 + z_{eq})
\int_{\eta_i}^{\eta_0} \, d\eta\, T_c(k;\eta) 
\widetilde\Theta_{+}(k;\eta)\, ,
\end{eqnarray}
where $\eta_i$ is the time when the network of cosmic defects
was generated and $T_c$ is the transfer function for the subsequent
perturbations, those
generated actively by the defects. In order to include 
compensation for the initial
perturbations, $\delta_m^I$, the substitution is usually made:
\bigbreak
\begin{equation} T_c(k;\eta) = \Big(1 + (k_c/k)^2 \Big)^{-1} \, T(k;\eta)\, ,
\label{compen}
\end{equation}
where $k_c$ is a long-wavelength cut-off at the compensation scale. In the 
case where the defects have several possible decay channels\footnote{For
the case of cosmic 
strings these could be loops, gravitational radiation and in our theory 
particles with velocity $c_p \gg c$.} there may be several scales 
associated with compensation due to the different dynamics of the 
defect decay products. However, for simplicity 
we shall assume that eqn. (\ref{compen}) is a good approximation 
with the compensation scale, $k_c$, determined by particles generated 
by the defect network with $c_p \gg c$. Consequently, we expect 
$k_c \sim  (c_p \eta)^{-1}$. If there are other decay channels this 
assumption may slightly alter the power spectrum normalization but 
will not otherwise affect the predictions of our model.

Note also that both in the present work and in our previous work
\cite{previous} we get a compensation scale that is much larger than
$c\eta$, but the reasons for this are slightly different in the two
cases. In \cite{previous} the compensation scale is expected to be of
the order of the correlation length of the network, while in the
present work the compensation scale is expected to be of the order of
the free-streaming length of the decay products.

For $(c_p \eta_0)^{-1}  \ll k \ll (c_p \eta_i)^{-1}$ the analytic
expression for the power spectrum of
density perturbations induced by defects can be written as 
\begin{equation}
P(k)=16 \pi^2 G^2 (1+z_{\rm eq})^2 \int_0^{\infty}d\eta
{\cal F}(k,\eta)|T_c(k,\eta)|^{2}\, ,
\label{pspec}
\end{equation}
where $ {\cal F}(k,\eta)$ is the structure function which can be obtained 
directly from the unequal time correlators \cite{WASA,AS}. 
Still in the case of super-horizon perturbations,
it can easily be shown \cite{WASA} that for a scaling 
network ${\cal F}(k,\eta)= {\cal F}(k \eta)$ which, combined with eqns. 
(\ref{T0}), (\ref{compen}) and (\ref{pspec}), gives 
\begin{equation}
P(k) \propto \int_0^{\infty}d\eta {\cal S}(k \eta) / \eta^2 \propto k\, 
\label{pspec2}
\end{equation}
for super-horizon modes. Here the function ${\cal S}$, is just the 
structure function, ${\cal F}$, 
times the compensation cut-off function.

Up until now we only considered the spectrum of primordial fluctuations 
induced by cosmic defects (by primordial we mean generated at very early 
times). In our model a Harrison-Zel'dovich spectrum is predicted (see eqn. 
(\ref{pspec2})) just as in the simplest inflationary models. The
final processed 
spectrum 
taking into account the growth of the perturbations inside the horizon in 
the radiation and matter eras will also be the same as for the simplest 
inflationary models. Note that if $c_p$ is a time varying quantity then 
the compensation cut-off is no longer a function of $k \eta$ alone, and so 
it is possible to have deviations from a pure Harrison-Zel'dovich spectrum 
just as in generic inflationary models.

On large scales $k \ll (c \eta)^{-1}$ the structure function 
${\cal F}(k, \eta)$ has a white noise spectrum. The turn-over scale, 
if it exists, only appears at the correlation length of the 
network $k_\xi \gsim (c \eta)^{-1}$ \cite{ASWA1,WASA}.
This means that perturbations induced on scales larger than the 
correlation length are generated by many defect elements and, therefore, 
have a Gaussian distribution according to the central limit theorem. 
On the other hand, perturbations induced on smaller scales are
very non-Gaussian because they can be either very large within the 
regions where a string has passed by or else very small outside these. 
This allows us to roughly divide the power spectrum 
of cosmic-string-seeded density perturbations 
into a nearly Gaussian component generated when the string 
correlation length was smaller than the scale under consideration,
and a strongly skewed non-Gaussian component generated 
when the string correlation length was larger (we call these 
the `Gaussian' and `non-Gaussian' contributions respectively).

The ratio of these two components may be easily computed by splitting 
the structure function in (\ref{pspec}), in two parts: a Gaussian part
 ${\cal F}_{\rm g}(k,\eta) = {\cal F}(k,\eta)$ for 
$k<k_\xi$ (${\cal F}_{\rm g}=0$ for $k>k_\xi$) and a non-Gaussian part 
${\cal F}_{\rm ng}(k,\eta)= {\cal F}(k,\eta)$ for $k>k_\xi$ 
(${\cal F}_{\rm ng}=0$ for $k<k_\xi$).
We can then integrate (\ref{pspec}) with this Gaussian/non-Gaussian split, 
to compute the relative contributions to the total power spectrum. The final 
result will of course depend on the choice of compensation scale $k_c$,
but recall that in any case we expect the network correlation length to
be much smaller than the compensation scale. Thus,
given that in our model $k_c \ll (c \eta)^{-1}$ the `non-Gaussian' 
component will simply be too weak to be detected. 

By allowing for a characteristic velocity for one of the decay 
channels of a defect network $c_p$ 
much larger than the velocity of light (and gravity) $c$ we were able to 
construct a model with primordial, adiabatic ($\delta_r=4 \delta_m/3$), 
nearly Gaussian fluctuations whose primordial spectrum is of the 
Harrison-Zel'dovich form. This model is indistinguishable from 
the simplest inflationary models (as far as structure formation is concerned). The $C_l$ spectrum and the polarization curves 
of the CMBR predicted by this model should also be identical to the ones 
predicted in the simplest inflationary models as the perturbations in the 
CMB are not generated directly by the defects. Moreover, the gravitational wave 
background generated by the defects in this theory should be too weak to be 
detected as the energy scale of the defects can be significantly lower than 
in the standard case.

\section{Discussion and conclusions}
\label{concl}

Following up our previous work on the cosmological consequences
of ``varying speed of light''
theories \cite{AM2,AMR,previous}, we have presented a further general
illustration of
non-negligible overlap between topological defect and inflationary
structure formation models, in the context of ``two-metric'' theories of
gravity.

According to Liddle's criterion \cite{hor}, such ``mimic'' models of
inflation  require some form of ``violation of causality''. 
In our previous work \cite{previous},  this
was provided by a defect-producing scalar field with
a characteristic speed much larger than that of gravity and the standard
model interactions. In the present work, it is instead provided by a
similarly large ``superluminal'' characteristic speed of the decay products
of the defect network.
Hence we see that defects in two-metric theories can produce either Gaussian
or non-Gaussian \cite{previous} primordial fluctuations.
The only distinguishing characteristics of these models, by comparison with
the simplest inflationary models, will be a small non-Gaussian signal in the
former case, and a background of the defect decay products in the latter.
Both of these could be detected by future experiments.
A more detailed discussion will be presented in \cite{fut}.

As we already pointed out in \cite{previous}, these models might admittedly
seem somewhat ``unnatural'' in the context of our present
theoretical prejudices, though they are certainly not the only ones to fit
in this category \cite{turok,desi}. However, if one keeps in mind
that any fully consistent cosmological structure formation model candidate
should eventually be derivable from fundamental physics, one could argue that
at this stage they are, {\em caeteris paribus}, on the same footing as
inflation. Certainly no single fully consistent realization of an
inflationary model is known at present.

In any case, what these models do provide is explicit evidence
of the fact that one must be extremely careful with one's prior assumptions
when using cosmological datasets, and that one must keep looking for
efficient and unambiguous ways to test the main paradigms of cosmology.
We shall return to this topic in a future publication.

\acknowledgements

We thank Paul Shellard for useful discussions and comments.
C.M. is funded by FCT (Portugal) under
`Programa PRAXIS XXI', grant no. PRAXIS XXI/BPD/11769/97.
We thank Centro de Astrof{\' \i}sica 
da Universidade do Porto (CAUP) for the facilities provided.

\end{document}